# Development of Focused X-ray Luminescence Compute Tomography Imaging

Yile Fang[1], Yibing Zhang[1], Changqing Li[2]*

[1]Department of Bioengineering, University of California, Merced, Merced, CA 95343, USA.
[2]Department of Electrical Engineering, University of California, Merced, Merced, CA 95343, USA.
*Corresponding Author: Changqing Li, Email: cli32@ucmerced.edu

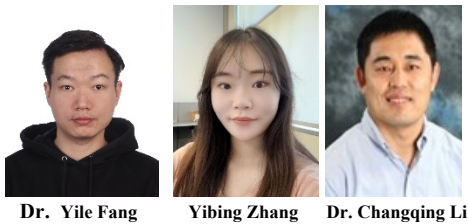

**Dr. Yile Fang**    **Yibing Zhang**    **Dr. Changqing Li**

X-ray luminescence is produced when contrast agents absorb energy from X-ray photons and release a portion of that energy by emitting photons in the visible and near-infrared range. X-ray luminescence computed tomography (XLCT) was introduced in the past decade as a hybrid molecular imaging modality combining the merits of both X-ray imaging (high spatial resolution) and optical imaging (high sensitivity to tracer nanophosphors).

**Narrow beam XLCT imaging**

Particularly, the narrow X-ray beam based XLCT has been demonstrated to have very high spatial resolution at depths of several centimeters with good molecular sensitivity inside turbid media [1, 2]. As shown in Fig. 1, X-ray photons excite X-ray excitable contrast agents emitting optical photons that propagate to the object surface to be measured by photodetectors. These measurements are used for model based XLCT image reconstructions usually with the anatomical guidance of X-ray beam position for better spatial resolution.

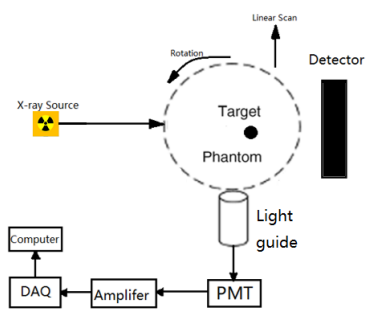

**Figure 1.** Schematic diagram of the narrow beam XLCT imaging system.

The first demonstration of XLCT imaging was reported by Pratx et al. [3] using a selective excitation scanning scheme, like the pencil beam X-ray CT imaging [3, 4] and then by many other groups [5-7]. We have shown that by using a focused beam of x-rays as the excitation source in XLCT, orders of magnitude of better sensitivity were achieved due to higher flux and efficient use of x-ray photons compared with the collimation-based method [8]. Our studies have demonstrated that the spatial resolution could be improved to be close to the X-ray beam size by reducing the scanning step size to be smaller than the X-ray beam size [9]. To perform multi-color XLCT imaging, we synthesized biocompatible nanophosphors with bright and distinct X-ray luminescence spectra and compared them with commercially available nanophosphors [10]. However, the long scan time of the narrow beam based XLCT limits its applications in 3D and in vivo imaging.

**Benchtop Focused XLCT imaging System**

Fig. 2 shows a benchtop focused beam based XLCT imaging system we have built. To improve the scan time, we have developed a continuous scanning scheme where the X-ray beam moves across the object continuously [11]. Furthermore, we have shown that the data acquisition time can be further reduced by using a gated photon counter to replace the high-speed oscilloscope [12]. With the proposed superfast scan scheme, we have achieved 43 seconds per transverse scan, which is 28.6 times faster than before with slightly better XLCT image quality [13]. The superfast scan allows us to perform 3D XLCT imaging and pencil beam CT imaging simultaneously.

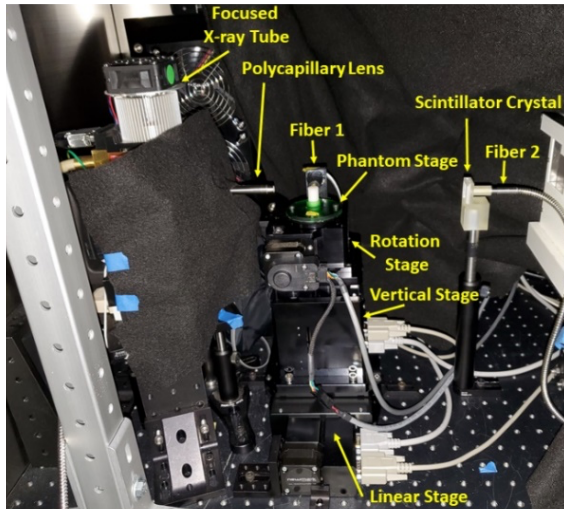

**Figure 2.** Photographs of the 3D XLCT imaging system.

**Focused beam based XLCT Imaging System with a Rotary Gantry**

So far, only benchtop systems have been developed to show proof-of-principle with phantom experiments [14, 15]. To perform in vivo imaging for small animals, we have designed and built a first-of-its-kind Focused X-ray Luminescence Tomography (FXLT) imaging system based on a rotary gantry as shown in Fig. 3 and Fig. 4 [16]. There is a co-registered cone beam microCT imaging system using a cone beam x-ray tube. We are able to perform both XLCT imaging and a pencil beam based microCT simultaneously using the superfine focused beam x-ray tube (fleX-Beam, XOS). All the major devices are mounted on the rotary gantry while the imaged objects such as small animals are placed on a linear stage in the rotary center.

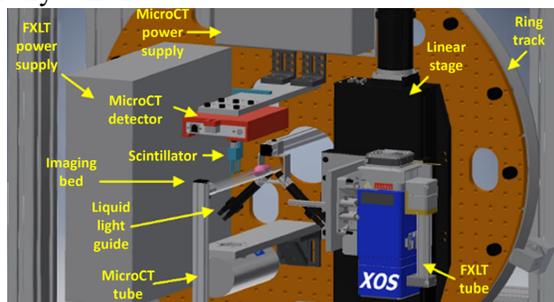

**Figure 3.** Design of the FXLT imaging system.

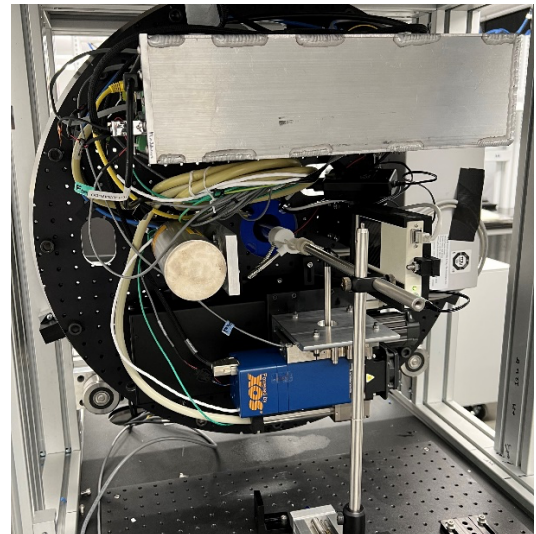

**Figure 4.** Photographs of the FXLT imaging system.

A lab-made C++ program was coded to control the imaging system. The linear scan at each angular projection was performed by moving the focused X-ray tube with the heavy-duty linear stage in a continuous motion using a fly-scanning scheme [13]. Using synthesized biocompatible nanophosphors as contrast agent, we have performed phantom experiments with capillary tube targets and 3D printed targets, scanned both euthanized and in vivo mice with xenografted tumors to evaluate the performance of the imaging system. The results indicated that the scanner is able to obtain in vivo 3D functional images and 3D structural images simultaneously, at high spatial resolution (150 µm), and with good molecular sensitivity (around 0.18 µM) in deep tissue. The nanoparticles inside the xenografted tumors of a live mouse have been successfully reconstructed three-dimensionally by the pencil beam XLCT for the first time.

**Future Directions**

In future studies, we will use the proposed XLCT imaging modality to image the molecular oxygen with high spatial resolution [14]. We will also use the scan scheme and reconstruction algorithm in XLCT to perform X-ray fluorescence computed tomography imaging with the focused beam x-ray tube [19].